\documentclass[useAMS,usenatbib,nofootinbib]{mn2e}
\usepackage{lmodern}
\usepackage[T1]{fontenc}
\usepackage[latin9]{inputenc}
\usepackage{amsmath,amssymb}
\usepackage[pdftex]{graphicx}

\usepackage{hyperref}
\usepackage[usenames,dvipsnames]{color}
\definecolor{menublue}{rgb}{0.0,0.0,0.5}
\definecolor{citegreen}{rgb}{0.0,0.5,0.5}
\definecolor{urlred}{rgb}{1.0,0.0,0.0}
\hypersetup{bookmarksopen,pdfstartview={FitH},colorlinks=true, breaklinks=true,menucolor=menublue,urlcolor=urlred,citecolor=citegreen,linkcolor=blue}
\usepackage[all]{hypcap}
\usepackage{wasysym}

\usepackage{txfonts}

\def\del#1{{}}

\newcommand{\ltsima}{$\; \buildrel < \over \sim \;$}
\newcommand{\lsim}{\lower.5ex\hbox{\ltsima}}
\newcommand{\gtsima}{$\; \buildrel > \over \sim \;$}
\newcommand{\gsim}{\lower.5ex\hbox{\gtsima}}
\newcommand{\bra}{\langle}
\newcommand{\ket}{\rangle}

\newcommand{\dd}{\mathrm{d}}

\newcommand{\covnoell}{\hat{C}_\psi}

\newcommand{\invcovnoell}{\hat{C}^{-1}_\psi}

\newcommand{\trace}{\mathrm{tr}}

\newcommand{\chip}{{\chi^\prime}}

\newcommand{\boX}{\boldsymbol{d}}

\newcommand{\lprime}{\ell^\prime}
\newcommand{\mprime}{m^\prime}

\newcommand{\likeli}{\mathcal{L}}
\newcommand{\dalilink}{\href{http://lnasellentin.github.io/DALI/}{DALI}}

\newcommand{\dalipaper}{SAQ2014}

\title[Non-Gaussian likelihoods in weak-lensing]
{Non-Gaussian forecasts of weak lensing with and without priors}
\author[E. Sellentin, B.M. Sch{\"a}fer]
{Elena Sellentin$^1$\thanks{e-mail: sellentin@stud.uni-heidelberg.de} and Bj{\"o}rn Malte Sch\"afer$^2$\\
$^1$Institut f{\"ur} Theoretische Physik, Universit{\"a}t Heidelberg, Philosophenweg 16, 69120 Heidelberg\\
$^2$Astronomisches Recheninstitut, Zentrum f{\"u}r Astronomie der Universit{\"a}t Heidelberg, Philosophenweg 12, 69120 Heidelberg, Germany}

\begin{document}
\pagerange{\pageref{firstpage}--\pageref{lastpage}}
\pubyear{2015}
\maketitle
\label{firstpage}

% --- abstract --- %
\begin{abstract}
Assuming a Euclid-like weak lensing data set, we compare different methods of dealing with its inherent parameter degeneracies.
Including priors into a data analysis can mask the information content of a given data set alone. However, since the information content of a data set is usually estimated with the Fisher matrix, priors are added in order to enforce an approximately Gaussian likelihood. Here, we compare priorless forecasts to more conventional forecasts that use priors. We find strongly non-Gaussian likelihoods for 2d-weak lensing if no priors are used, which we approximate with the DALI-expansion. Without priors, the Fisher matrix  of the 2d-weak lensing likelihood includes unphysical values of $\Omega_m$ and $h$, since it does not capture the shape of the likelihood well. The Cramer-Rao inequality then does not need to apply. We find that DALI and Monte Carlo Markov Chains predict the presence of a dark energy with high significance, whereas a Fisher forecast of the same data set also allows decelerated expansion. We also find that a 2d-weak lensing analysis provides a sharp lower limit on the Hubble constant of $h > 0.4$, 
even if the equation of state of dark energy is jointly constrained by the data. This is not predicted by the Fisher matrix and usually masked in other works by  a sharp prior on $h$. Additionally, we find that DALI estimates Figures of Merit in the presence of non-Gaussianities better than the Fisher matrix.
We additionally demonstrate how DALI allows switching to a Hamiltonian Monte Carlo sampling of a highly curved likelihood with acceptance rates of $\approx 0.5$, an effective covering of the parameter space, and numerically effectively costless leapfrog steps. This shows how quick forecasts can be upgraded to accurate forecasts whenever needed. Results were gained with the public code from \dalilink . 
\end{abstract}

% --- keywords --- %

% --- section: introduction --- %
\section{Introduction}
Weak cosmic lensing is currently a field of intense focus: It allows the measurement of the cosmological parameters especially in the late Universe and is therefore an ideal probe for dark energy physics or models of modified gravity. After the first significant detections \citep{2000astro.ph..3338K,2000Natur.405..143W,2000MNRAS.318..625B,2000A&A...358...30V}, weak gravitational lensing has been observed with increasing singificance by e.g. CFHTLenS \citep{2013MNRAS.430.2200K, 2013MNRAS.432.2433H}, allowing the determination of cosmological parameters.

In the future, weak lensing will be investigated on about a third of the sky with the upcoming Euclid satellite \citep{EuclidStudyReport}. While the Euclid data set is not yet available, its constraining power on different extensions of the current cosmological standard model is being forecasted, see e.g. \citet{Lucabook}. Also, statistical techniques are being improved, or the data analysis is being refined, for example by switching from a two dimensional weak lensing analysis to weak lensing tomography \citep{1999ApJ...522L..21H, 2002PhRvD..66h3515H} and 3d weak cosmic shear \citep{2003MNRAS.343.1327H, 2005PhRvD..72b3516C,2006MNRAS.373..105H}, or by including higher-order polyspectra of the weak lensing shear \citep{2010arXiv1003.5003M}, or by combining lensing with other tracers of cosmological structure growth. There will be large-scale lensing surveys on the way to Euclid with an emphasis on dark energy, for instance the Kilo-degree Survey (KidS) \citep{2013Msngr.154...44J} and the Dark-Energy-survey (DES) \citep{2015MNRAS.449.2219M}.

All these different methods need a tool in order to asess the information content of the data set under a specific analysis. Usually, the wish is to quickly forecast the resulting likelihood constraints or Figure of Merits, and sometimes also Bayesian evidences. In principle, Monte Carlo Markov Chains (MCMC), Nested Sampling or grid-based likelihood evaluations are a well suited tool for these aims, but they are very time consuming. Quick estimates of the above quantities are then usually done with the Fisher matrix approach \citep{1997ApJ...480...22T} which hinges on the assumption of the likelihood being well approximated by a multivariate Gaussian. However, a Gaussian likelihood can only be gained under additional assumptions about the data set and the parametric model that is fitted to the data:

The Gaussian shape of the likelihood will only be achieved if the signal depends linearly on the model parameters. If the data depends non-linearly on parameters, this non-linearity can cause the likelihood to be non-Gaussian if the data set is not well-constraining such that a linear Taylor approximation around the maximum likelihood point is enough to capture the variation of the physical model within the parameter space that is preferred by the data. Particularly severe non-Gaussianities can be expected if non-linear parameters are in addition strongly degenerate with each other.

Such non-Gaussianities lead to a broad variety of likelihood contour shapes, amongst which banana-shapes are often observed, as well as straight but asymmetric contours. Such asymmetries can be for example introduced, if the best fit point lies close to the boundary of an unphysical region.

These non-ellipsoidal and non-symmetric likelihood shapes are not captured by the Fisher matrix analysis. The Fisher matrix might then result in confidence contours that easily extend beyond the physically meaningful parameter range. The problem can worsen if parameters are marginalized over: In cases where the Fisher matrix captures the orientation of the likelihood wrongly along one parameter direction, all other parameters will be affected by this, since they are being constrained jointly. If the Fisher matrix extents into unphysical regions, the Cramer-Rao bound also does not need to be fulfilled, since it holds under the condition that the Fisher information is defined and finite everywhere within the covered data and parameter space. We discuss this issue in Sect.~\ref{sect_comparison} for the example of 2d weak lensing.

All these problems are traditionally addressed by combining likelihoods from different probes, or by imposing priors such that parameter degeneracies are broken and the combined likelihood is more sharply peaked and therefore confined to the physically meaningful parameter space. This solves the above problems by removing non-Gaussianities. Another solution would be to accurately capture existing non-Gaussianities. The latter is possible with the DALI-approach \citep[][henceforth ~\dalipaper]{Sellentinetal}\citep{Sellentin2015}, which we shall in the following compare to the Fisher matrix approach and to MCMC-evaluated likelihoods.

We adopt the Einstein sum convention for repeated indices. Our cosmological parameter set consists of $\theta = (\Omega_m, \sigma_8, n_s, h, w)$ which are the density of cold dark matter today, the normalization of the power spectrum, the primordial spectral index, the Hubble constant and a redshift independent dark energy equation of state parameter. Our fiducial cosmology is ${\Omega_m = 0.25, \sigma_8 =0.8, n_s=0.96, h=0.7, w=-0.98}$. We keep the density of baryons fixed to $\Omega_b = 0.04$.

This paper is organized as follows: we describe the modeling of the weak lensing observations and why degeneracies can be expected in Sect.~\ref{sect_cosmology}. The likelihood and its approximations are described in Sects.~\ref{sect_statistics} and~\ref{sect_approx}. Sect.~\ref{sect_comparison} contains a comparison of the Fisher matrix, DALI and MCMC, and describes the advantages of accounting for non-Gaussian degeneracies instead of removing them by the use of priors. We use DALI as approximate potential for a Hamilton Monte Carlo Sampler in Sect.~\ref{sect_mcmc}. Sect.~\ref{sect_summary} presents a summary of our results.

% --- section: cosmology ---%
\section{cosmology and weak lensing}\label{sect_cosmology}
In spatially flat dark energy cosmologies with redshift independent equation of state parameter $w$, one obtains for the Hubble function $H(a)=\dd\ln a/\dd t$,
\begin{equation}
\frac{H^2(a)}{H_0^2} = \frac{\Omega_m}{a^{3}} + \frac{1-\Omega_m}{a^{3(1+w)}}.
\end{equation}
The comoving distance $\chi$ is related to the scale factor $a$ through
\begin{equation}
\chi = -c\int_1^a\:\frac{\dd a}{a^2 H(a)},
\end{equation}
where the Hubble distance $\chi_H=c/H_0$ is the natural unit for cosmological distance measures. Small fluctuations $\delta$ in the distribution of cold dark matter grow in the linear regime $\left|\delta\right|\ll 1$ \citep{2003MNRAS.346..573L} according to
\begin{equation}
\frac{\dd^2}{\dd a^2}\delta(a) +
\frac{1}{a}\left(3+\frac{\dd\ln H}{\dd\ln a}\right)\frac{\dd}{\dd a}\delta(a) -
\frac{3}{2a^2}\Omega_m(a) \delta(a) = 0,
\label{eqn_growth}
\end{equation}
and their statistics is characterised by the spectrum $\bra \delta(\bmath{k})\delta(\bmath{k}^\prime)\ket = (2\pi)^3\delta_D(\bmath{k}+\bmath{k}^\prime)P_\delta(k)$ with the ansatz $P_\delta(k)\propto k^{n_s}T^2(k)$ using the transfer function $T(k)$, while they grow proportionally to the growth function $D_+(a) = \delta(a)/\delta(1)$. The transfer function depends on the Hubble constant through the shape parameter $\Gamma = \Omega_mh$ \citep{1986ApJ...304...15B, 1995ApJS..100..281S}. The spectrum is normalised to the value $\sigma_8$,
\begin{equation}
\sigma_8^2 = \int_0^\infty\frac{k^2\dd k}{2\pi^2}\: W^2(8~\mathrm{Mpc}/h\times k)\:P_\delta(k),
\end{equation}
with a Fourier-transformed spherical top-hat $W(x) = 3j_1(x)/x$ as the filter function, where $j_1(x)$ is the spherical Bessel function of the first kind. From the CDM spectrum of the density perturbations the spectrum of the dimensionless Newtonian gravitational potential $\Phi$ can be obtained
\begin{equation}
P_\Phi(k) \propto \left(\frac{3\Omega_m}{2\chi_H^2}\right)^2\:k^{n_s-4}\:T(k)^2,
\end{equation}
by applying the comoving Poisson-equation $\Delta\Phi = 3\Omega_m/(2\chi_H^2)\delta$ for converting between density contrast $\delta$ and gravitational potential $\Phi$. Additional variance of the cosmic density field on nonlinear scales is described by \citet{2003MNRAS.341.1311S}, which we include in our modelling.

Weak gravitational lensing probes the tidal gravitational fields of the cosmic large-scale structure by the distortion of light bundles \citep{2001PhR...340..291B,2010CQGra..27w3001B}. To make best use of the cosmological information, one divides the galaxy sample from which shape correlation functions or spectra are estimated \citep{2004ApJ...601L...1T, 2007MNRAS.381.1018A, 2005PhRvD..72d3002H, 2003PhRvL..91n1302J, 2004MNRAS.348..897T, 2012MNRAS.423.3445S}, into $n_\mathrm{bin}$ redshift intervals and computes the lensing potential $\psi$ at the position $\bmath{\theta}$ for each redshift bin $i$ separately,
\begin{equation}
\psi_i(\bmath{\theta}) = \int_0^{\chi_H}\dd\chi\:W_i(\chi)\Phi,
\label{eqn_lensing_potential}
\end{equation}
hence $\psi_i(\bmath{\theta})$ is related to the gravitational potential $\Phi$ by projection with the weight function $W_i(\chi)$ which contains physical information since it depends depends on background geometry and also on the growth of matter as
\begin{equation}
W_i(\chi) = 2\frac{D_+(a)}{a}\frac{G_i(\chi)}{\chi}.
\end{equation}
Modes $\psi_{\ell m,i}$ of the lensing potential $\psi_i(\bmath{\theta})$ are obtained by the decomposition $\psi_{\ell m,i} = \int\dd\Omega\:\psi_i(\bmath{\theta})Y_{\ell m}^*(\bmath{\theta})$ into spherical harmonics $Y_{\ell m}(\bmath{\theta})$. The distribution $p(z)\dd z$ of the lensed galaxies in redshift is incorporated in the lensing efficiency function $G_i(\chi)$,
\begin{equation}
G_i(\chi) = \int^{\chi_{i+1}}_{\mathrm{min}(\chi,\chi_i)}\dd\chip\: 
p(\chip) \frac{\dd z}{\dd\chip}\left(1-\frac{\chi}{\chip}\right)
\end{equation}
with $\dd z/\dd\chip = H(\chip) / c$ and the bin edges $\chi_i$ and $\chi_{i+1}$, respectively. Euclid forecasts commonly use the parameterisation \citep{2008arXiv0802.2522R}
\begin{equation}
p(z)\dd z \propto \left(\frac{z}{z_0}\right)^2\exp\left(-\left(\frac{z}{z_0}\right)^\beta\right)\dd z.
\end{equation}
Angular spectra $C_{\psi,ij}(\ell)$ of the tomographic weak lensing potential can be written as the variance $\bra\psi_{\ell m,i}\psi_{\lprime\mprime,j}^*\ket=\delta_{\ell\lprime}\delta_{m\mprime} C_{ij}(\ell)$, which we approximate by the corresponding flat-sky expression,
\begin{equation}
C_{\psi,ij}(\ell) = \int_0^{\chi_H}\frac{\dd\chi}{\chi^2}\:W_i(\chi)W_j(\chi)\:P_\Phi(k=\ell/\chi).
\end{equation}
The convergence $\kappa$ and the shear $\gamma$ follow by double differentiation of the lensing potential with respect to angles, $\kappa=\ell^2\psi/2$, therefore their spectra are equal to $\ell^4C_{\psi,ij}(\ell)/4$. Observed spectra of the weak lensing shear will contain a constant contribution $\sigma_\epsilon^2 n_\mathrm{bin}/\bar{n}$ known as shape noise, which translates into
\begin{equation}
\hat{C}_{\psi,ij}(\ell) = C_{\psi,ij}(\ell) + \sigma_\epsilon^2\frac{n_\mathrm{bin}}{\bar{n}}\times\ell^4\:\delta_{ij},
\end{equation}
which will be at the same time the covariance matrix for measurements of modes $\psi_{\ell m,i}$. $C_{\psi,ij}(\ell)$ is non-zero for redshift bins $i\neq j$ because light rays share the section between the observer and the closer tomography bin, such that they contain partially the same statistical information, leading to a non-vanishing covariance. We will mainly work with 2d-weak lensing, but for an additional comparison with 2-bin tomography, we choose bins such that they contain the same number $\bar{n}$ of galaxies.

\subsection{Curved degeneracy lines in weak lensing}
Weak gravitational lensing derives its sensitivity on cosmological parameters from a combination of the amplitudes of gravitational potentials and geometry: Directly, it depends on $(\Omega_m\sigma_8)^2$ because this sets the strength of the gravitational potential. Therefore, the hyperbolic degeneracy line between these two parameters will follow $\Omega_m \propto \sigma_8^{-1}$. The gravitational potentials grow with $D_+(a)/a$, which depends also on the dark energy equation of state parameter $w$. The dark energy equation of state also influences the conversion between distance measures and redshift because the expansion of the Universe is modified. More negative $w$ makes comoving distances for a given redshift larger, thus increasing the lensing signal. The shape of the spectrum $P(k)$ is determined by $n_s$, $h$ and $\Omega_m$, which is reflected in the spectra $C_{\psi,ij}(\ell)$ more weakly due to the weighting functions $W_i(\chi)$. The shape parameter of the power 
spectrum is $\Gamma = \Omega_m h$, which introduces a degeneracy between $\Omega_m$ and $h$ which will again be a hyperbolic line.

\begin{figure}
\centering \includegraphics[width=0.45\textwidth]{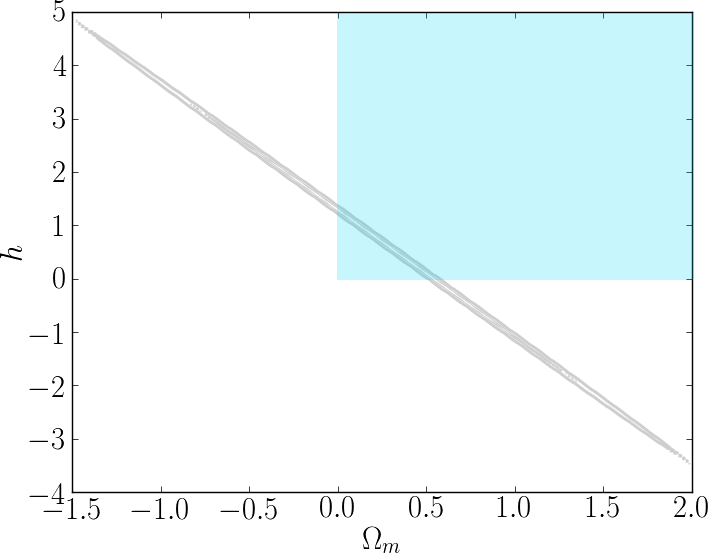} 
\caption{The marginalized Fisher matrix (grey) in the $\Omega_m,h$-plane. The blue rectangle indicates the area bounded by the constraints $h > 0$, $\Omega_m > 0$ which might be interpreted as minimal priors that could be applied to foster the constraining power of the weak lensing data set.}
\label{Fish}
\end{figure}

% --- section: statistics--- %
\section{The unapproximated likelihood}\label{sect_statistics}
for statistically homogeneous random fields, weak gravitational lensing yields independent modes in multipole $\ell$ and in the case of statistical isotropy, a measurement of $2\ell+1$ independent modes for each multipole. The likelihood for a model $\hat{C}_\psi(\ell)$ to able to reproduce the set $\left\{\psi_{\ell m,i}\right\}$ of observed modes $\psi_{\ell m,i}$ separates in the ideal case in $\ell$ and $m$ according to
\begin{equation}
\likeli\left(\left\{\psi_{\ell m,i}\right\}\right) = \prod_\ell \likeli\left(\psi_{\ell,i}|\hat{C}_{\psi,ij}(\ell)\right)^{2\ell + 1},
\label{eqn_lensing_likely}
\end{equation}
where the equal likelihoods of all modes $m$ at fixed $\ell$ have been multiplied. 

Observational issues like an incomplete sky coverage and a point spread function lead to a coupling of different $\ell$. Although of course relevant for analyses of real data sets, this is often omitted for forecasting \citep{2006JCAP...06..025H} because non-diagonal or non-block diagonal covariance matrices are  much harder to invert. For a Monte Carlo sampler, this inversion is necessary for each sample, and we therefore assume that the different $\ell$-modes decouple for the sake of speed.

The likelihood for each observed mode $\psi_{\ell m,i}$ if the theory predicts a covariance $\hat{C}_{\psi,ij}(\ell)$, is Gaussian in the data,
\begin{equation}
\likeli\left(\psi_{\ell m,i}\right) = 
\frac{1}{\sqrt{(2\pi)^{n_\mathrm{bin}}\mathrm{det}\hat{C}_\psi(\ell)}}\exp\left(-\frac{1}{2}\psi_{\ell m,i}(\hat{C}_{\psi}(\ell)^{-1})_{ij}\psi_{\ell m,j}\right),
\end{equation}
due the fact that both the cosmic structures as well as the noise are approximately Gaussian random fields. Consequently, the logarithmic likelihood $L = -\ln\likeli$ is up to an additive constant equal to 
\begin{equation}
L = \sum_\ell\frac{2\ell+1}{2}\:\left(\trace \ln\hat{C}_{\psi} + (\hat{C}_\psi^{-1})_{ij}\:\psi_{\ell m,i}\psi_{\ell m,j}\right)
\label{MCMC-likelihood}
\end{equation}
by using the relation $\mathrm{det}\ln C = \trace\ln C$ for the matrix $C_{\psi,ij}(\ell)$, indexed by the tomography bin numbers. We will often refer to Eq.~(\ref{MCMC-likelihood}) as the true likelihood since no approximations apart from physical approximations such as the flat sky approximation and integrating the lensing signal along a straight line were used so far, and the assumption that the projected lensing potential has nearly Gaussian fluctuation statistics.

We model Euclid's weak lensing survey \citep{EuclidStudyReport} to reach out to a median redshift of $0.9$ and to yield $\bar{n}=4.8\times10^8$ galaxies per steradian. We assume that the shape measurement produces a Gaussian noise with standard deviation $\sigma_\epsilon=0.3$. We use a sky fraction of $f_\mathrm{sky} = 0.35$ and a multipole range of $30$ to $3000$.

\begin{figure*}
\centering \includegraphics[width=\textwidth]{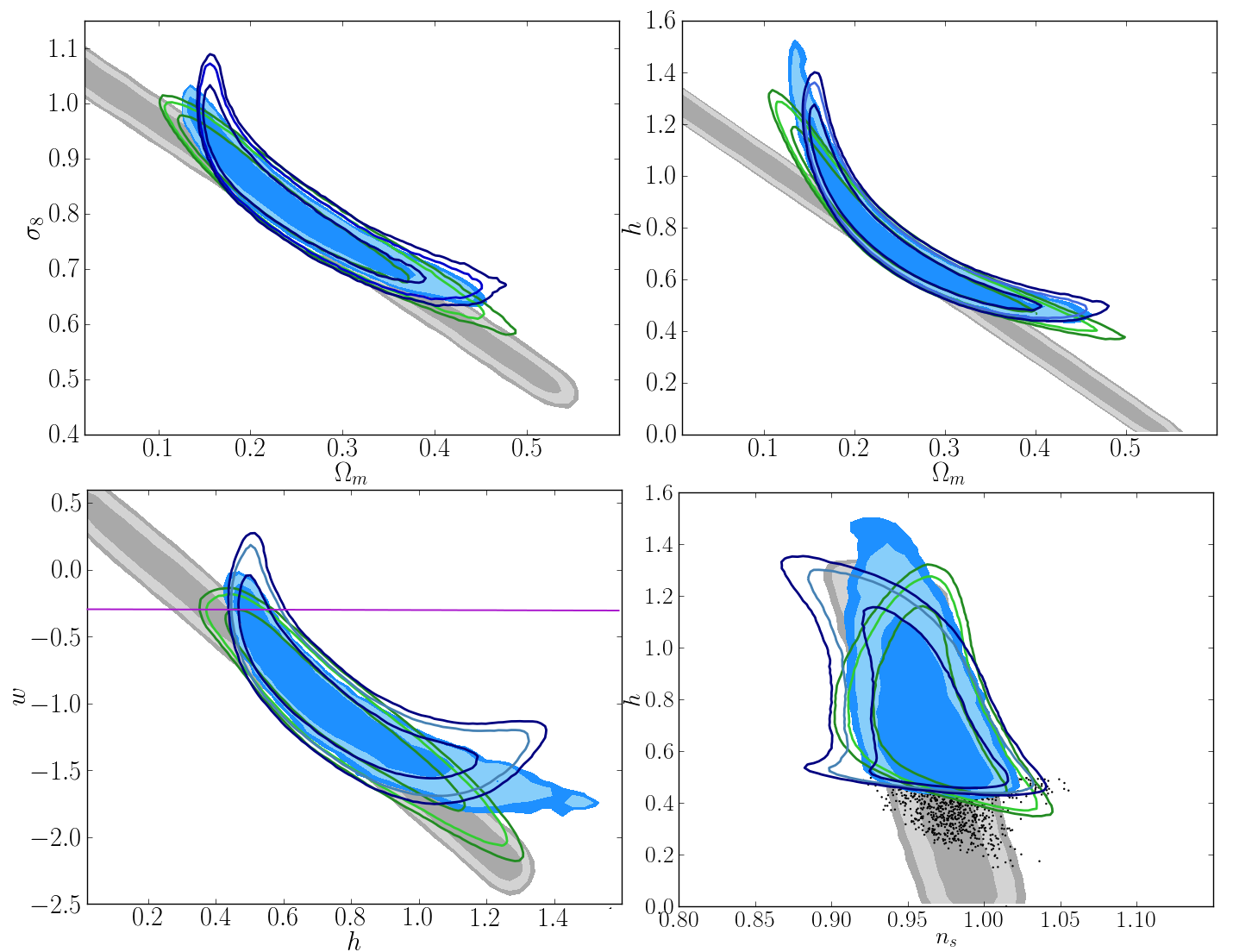} 
\caption{Comparison of the different likelihood approximations with the MCMC-sampled likelihood for 2d-weak lensing. The contours enclose $68\%, 95\%$ and $99\%$ of the likelihood. Solid grey: Fisher approximation combined with the additional constraints $\Omega_m > 0$ and $h > 0$ (implemented by sampling from the Fisher matrix and discarding all unphysical samples before marginalizing). DALI with second-order derivatives of the covariance matrix is shown in open green contours, DALI with second and third derivatives is shown in open blue contours. The dots in the bottom right panel are samples drawn from the Fisher matrix approximated likelihood and are predicted to be points of high likelihood by the Fisher matrix. However, when calculating the unapproximated likelihood of these samples, they turn out to be extremely unlikely parameter combinations. This demonstrates that the sharp cutoff towards lower $h$ in the bottom right panel is correct. In the top left and bottom right panel, the likelihood 
asymptotes roughly towards $h \approx 0.4$, and the cutoff in the $n_s,h$-plane is just a different projection of this behaviour. The purple line in the bottom left panel indicates the constraint $w < -1/3$ for accelerated expansion, and it can be seen that over $\sim 90 \%$ of the MCMC and DALI contours fall within the parameter space of accelerated expansion, thereby indicating strongly the presence of a dark energy, whereas $\sim 30 \%$ of the Fisher matrix cover parameter regions that would not lead to accelerated expansion.}
\label{All6}
\end{figure*}

% ---  --- %
\section{The different likelihood approximations}\label{sect_approx}
The Fisher matrix approach and DALI use derivatives at the best fit point to approximate Eq.~(\ref{MCMC-likelihood}) with increasing precision: From the data-averaged curvature $\bra\partial_\mu\partial_\nu L\ket$ of the logarithmic likelihood one derives the Fisher matrix $F_{\mu\nu}$ \citep{1997ApJ...480...22T, 2003MNRAS.343.1327H},
\begin{equation}
F_{\mu\nu} = \sum_\ell\frac{2\ell+1}{2}\left(
\partial_\mu\ln\hat{C}_{\psi,ij}(\ell)\:\partial_\nu\ln\hat{C}_{\psi,ji}(\ell)\right)
\end{equation}
with $\partial_\mu$ being the derivatives with respect to individual cosmological parameters, $x_\mu = \left(\Omega_m,\sigma_8,n_s, h, w\right)$. 
The Fisher matrix $F_{\mu\nu}$ allows the construction of a Gaussian likelihood of inferred parameters,
\begin{equation}
p(x_\mu) \propto \exp\left(-\frac{1}{2}\Delta x_\mu F_{\mu\nu} \Delta x_\nu\right)
\end{equation}
with the distances $\Delta x_\mu$ of the parameters from the best fit point. 

DALI then introduces higher order derivatives of the likelihood in order to recover non-Gaussianities. To fourth order in $\Delta x_\mu$ the DALI-approximated likelihood reads 
\begin{equation}
\begin{aligned}
& \ln p(x_\mu) = \\
&  -\frac{1}{4} \sum_\ell(2\ell+1)\:\trace\left(
 \invcovnoell \partial_\mu\covnoell \invcovnoell \partial_\nu\covnoell\right)   \Delta x_\mu \Delta x_\nu\\
&-\frac{1}{4} \sum_\ell(2\ell+1)\:\trace\left(
 \invcovnoell \partial_\mu \partial_\gamma\covnoell \invcovnoell \partial_\nu\covnoell\right)   \Delta x_\mu \Delta x_\nu \Delta x_\gamma \\
&-\frac{1}{16} \sum_\ell(2\ell+1)\:\trace\left(
 \invcovnoell \partial_\mu \partial_\gamma\covnoell \invcovnoell \partial_\nu \partial_\delta\covnoell\right)   \Delta x_\mu \Delta x_\nu \Delta x_\gamma\Delta x_\delta \\
& + \mathcal{O}(\Delta x^5),
\label{dalilike}
\end{aligned}
\end{equation}
see \citet{Sellentin2015}. Here, the index $\ell$ of the covariance matrix has been suppressed for brevity. From the second line of Eq.~(\ref{dalilike}) onwards, the higher order derivatives of the covariance matrix give access to the non-linearity of the parameters. DALI can also include higher-order derivatives of in principle arbitrary order. For example, a first estimate of  non-Gaussianities present in the model can be achieved by including second derivatives with DALI, a refined estimate can be achieved by including also third order derivatives. In order to make DALI fast, stopping at third order derivatives is advised but not mandatory.

\section{Forecasting with non-Gaussianities}\label{sect_comparison}
In the following, we compare which information about the preferred parameter space can be extracted from the weak lensing data set from Sect.~\ref{sect_cosmology} when analysed with the Fisher matrix, DALI and MCMC. Additionally, we compare the Figure of Merits (FoM) from the different approximations: The Fisher matrix allows a convenient definition of a FoM via the determinant of $2\times2$ submatrices
\begin{equation}
 \mathrm{FoM} = \sqrt{\mathrm{det}\boldsymbol{F}_{2\times2}}.
\end{equation}
This corresponds to using the area enclosed by a chosen confidence contour in a given parameter plane as a FoM. We generalize this concept to our non-Gaussian forecasts by defining that the FoM shall be the area enclosed by the $95\%$-confidence contour.

We begin by evaluating the Fisher matrix for this setup. Fig.~\ref{Fish} shows the marginalized Fisher matrix approximated likelihood in the $\Omega_m,h$-plane. Clearly, the Fisher matrix reaches far into unphysical regions of negative $\Omega_m$. It also covers regions of negative Hubble constants. Sensitivity with respect to the Hubble constant enters weak lensing through the shape parameter $\Gamma = \Omega_m h$ of the power spectrum. However, the shape parameter is a length scale and must therefore be positive definite. Negative $h$ and $\Omega_m$ are therefore unsensical in the chosen parameterization of the power spectrum via a shape parameter, and these negative values must be excluded.

This shows, that the Fisher matrix cannot be used for a 2d weak lensing analysis for the Euclid satellite without enforcing by priors that the shape parameter has to be positive definite. In the appendix we discuss shortly why the Cramer-Rao inequality does not hold if unphysical parameter ranges are covered by the Fisher matrix. 

\begin{figure}
\centering \includegraphics[width=0.5\textwidth]{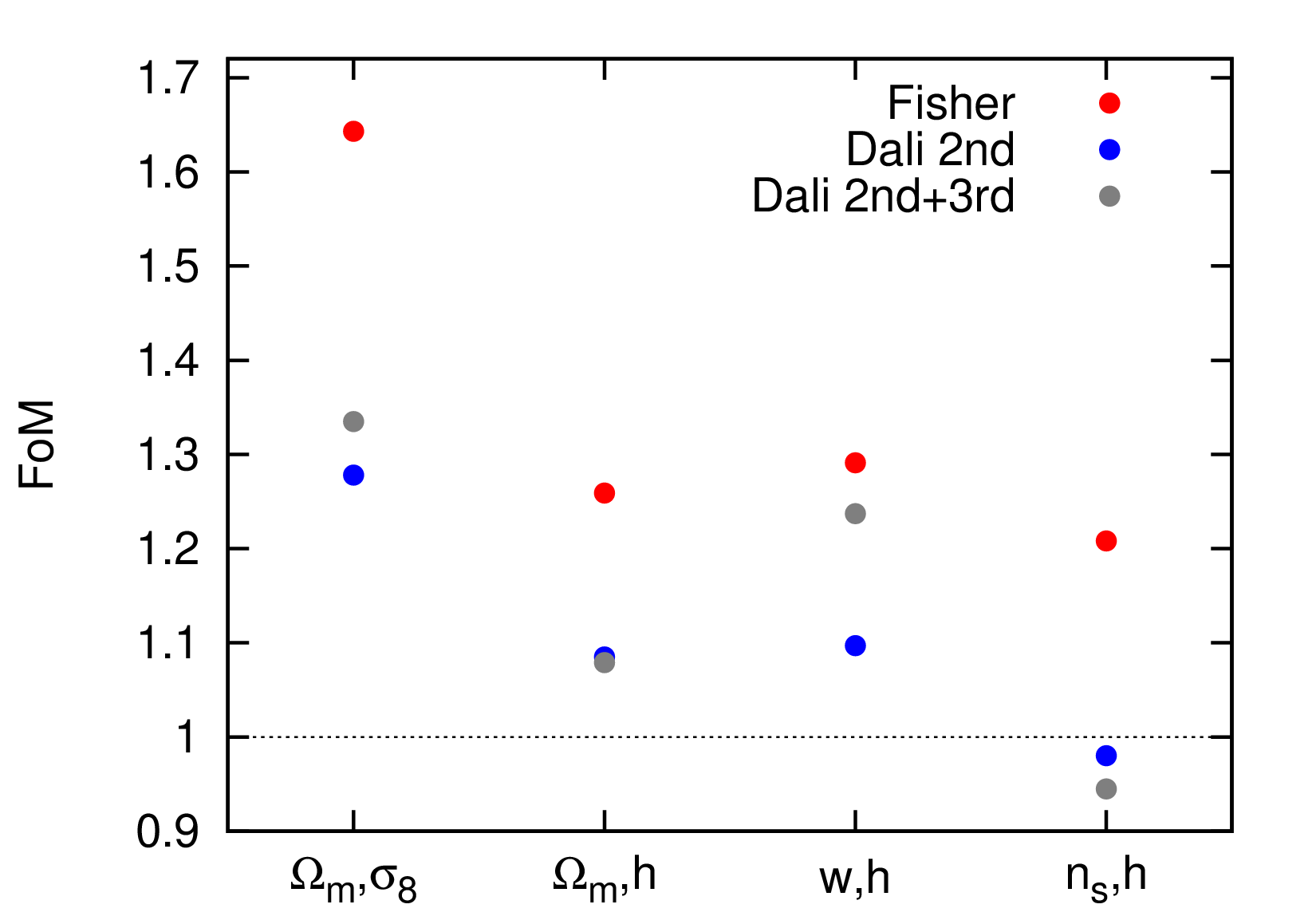} 
\caption{Figure of merit from the different approximations, relative to the MCMC figure of merit. The non-Gaussian DALI-approximations always perform better than the Fisher matrix, although no clear trend can be made out. However, the DALI-FoMs differ by maximally $\sim 30 \%$ from the MCMC-FoM, whereas the Fisher-FoM differs about two times more, namely by up to $ \sim 65 \%$.}
\label{FoM}
\end{figure}

For the comparison of the Fisher matrix with DALI and MCMC, we therefore augment the Fisher matrix with the prior knowledge $\Omega_m > 0, h > 0$. In practice, we implement this by drawing samples from the Fisher matrix approximated likelihood, and discarding all samples that fall into the unphysical regions. The introduction of these sharp cutoffs in $\Omega_m$ and $h$ leads to a non-Gaussian likelihood approximation. This approximation is depicted in grey in Fig.~\ref{All6} and was also used for comparing FoMs in Fig.~\ref{FoM}.

From a Fisher matrix analysis one would conclude that a 2d weak lensing analysis of a Euclid like survey does not allow to put a lower bound on the Hubble constant of our Universe. Additionally, the large uncertainty in $h$ and $\Omega_m$ leads to rather loose constraints of the remaining parameters $\sigma_8, n_s$ and $w$.

A comparison with MCMC-sampled likelihoods shows that the data are actually more constraining than predicted by the Fisher matrix, and a DALI-evaluation of the likelihood contours reveals that the problem is entirely due to non-Gaussianities and degeneracies between non-linear parameters. 

In Fig.~\ref{All6}, a comparison between the Fisher matrix, DALI and MCMC-samples of the likelihood is shown. For MCMC and DALI, no prior constraints like $\Omega_m > 0$ were used. Highly curved degeneracy lines and asymmetric likelihood shapes are evident. These curved degeneracy lines are well approximated by DALI, although not perfectly. As the likelihood asymptotes to $h \approx 0.4$ in the $\Omega_m,h$-plane and in the $h,w$-plane, negative $h$ are excluded without the use of any priors. This shows that the 2d weak lensing analysis is able to predict a lower bound of $h \geq 0.5$ on its own. Also due to the highly curved likelihood shapes, $\Omega_m$ does not become negative but stays confined to the physical region. These strong changes in the allowed range of $\Omega_m$ and $h$ in comparison to the Fisher matrix, propagate into the constraints of the remaining parameters $\sigma_8, n_s, w$. For dark energy, the curved DALI-approximation predicts $0.3 > w > -2.0$. In contrast, the Fisher 
matrix allows 
much smaller and even positive $w$. This is interesting for the forecasting of dark energy constraints: An accelerated expansion of the universe requires $w < -1/3$. About one third of the Fisher matrix covers however the parameter space $w > -1/3$, and only two thirds fall into the parameter range of accelerated expansion. In contrast, DALI and MCMC both favour the accelerated expansion by a much larger degree: about $90\%$ of their preferred parameter range corresponds to an accelerating universe.
Note, that the fact that the Fisher matrix also covers parameter regions of decelerated expansion stems from it being by construction symmetric around the best fit point. Also in \dalipaper ,~we observed that this high symmetry leads to the Fisher matrix covering parameter ranges of decelerated expansion, whereas the real likelihood did not.

The FoMs in the different parameter planes are compared in Fig.~\ref{FoM}, demonstrating that the DALI-FoMs are closer to the MCMC-FoM than the Fisher-FoM. 

In summary, the width of the Fisher ellipse perpendicular to the main degeneracy directions is in good agreement with the width of the true likelihood, whereas the semi-major axes parallel to the degeneracy lines are overestimated. DALI instead estimates well the constraining power of the data also along directions of strong degeneracies. This comparison shows, that the correct modelling of degeneracies between non-linear parameters can remove the necessity to break degeneracies by imposing priors in order to establish a Gaussian likelihood. The major advantage of modelling parameter degeneracies instead of breaking them with priors is that it allows to care only about a single data set, and to optimize its scientific return independently of external measurements. As a comparison, we shall in the following nonetheless show whether non-Gaussianities remain if a prior on $h$ is introduced or 2-bin tomography is performed.

\section{Breaking degeneracies}
Generally speaking, two ways of breaking degeneracies in a data set exist: Either, the data set is combined with external measurements, or the analysis of the data set is substantially improved. For weak lensing, the first option could be for example implemented by combining the data set with a prior on the Hubble constant. The second option could be implemented by changing from a 2d-analysis to tomographic weak lensing, since tomographic weak lensing is able to break degeneracies between cosmic parameters by the additional redshift information.

\begin{figure*}
\centering \includegraphics[width=\textwidth]{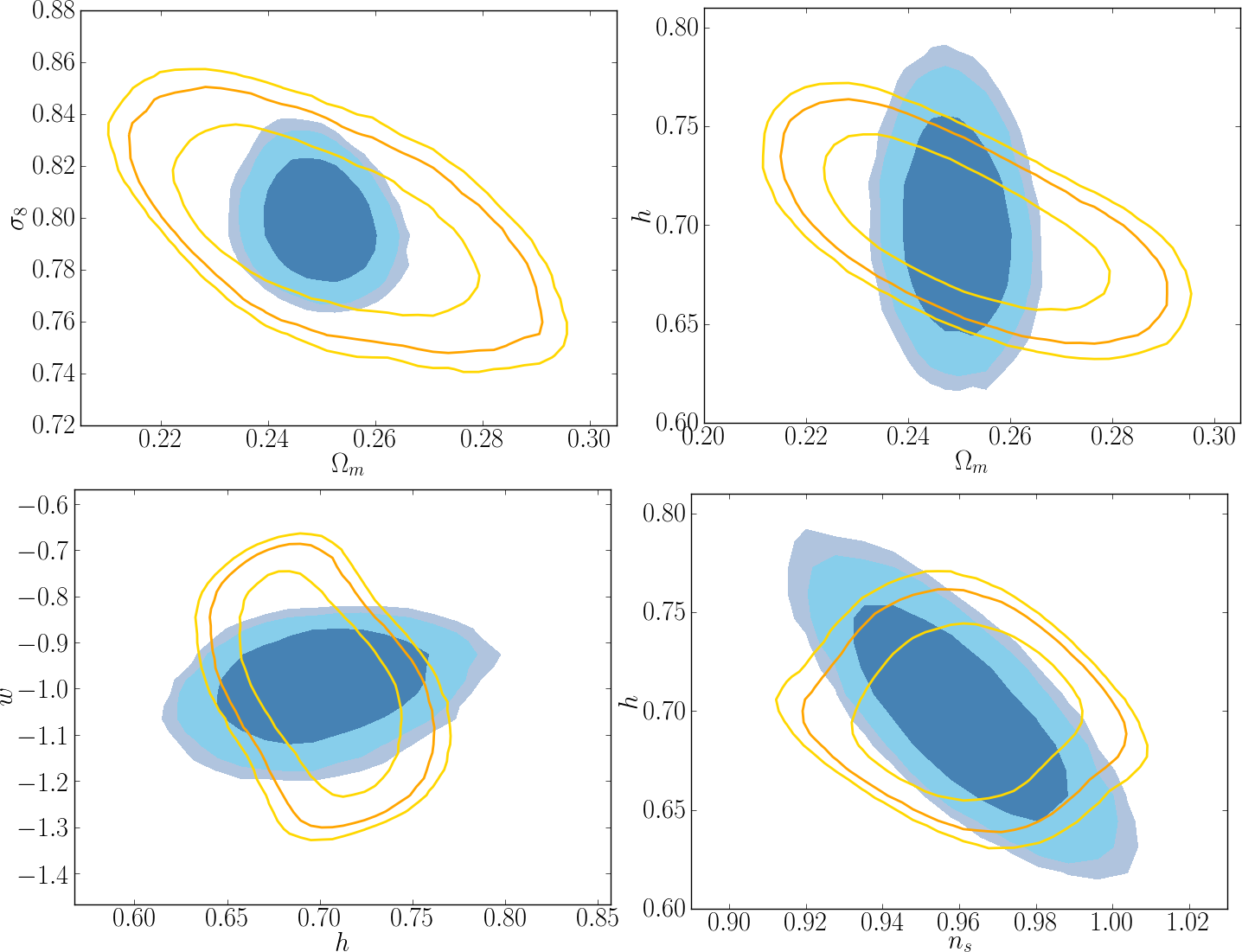} 
\caption{A comparison of 2d weak lensing and 2-bin tomography. Yellow: DALI-forecasted constraints for the 2d weak lensing analysis combined with a Gaussian prior on $h$ from local measurements. Since DALI with second order derivatives and DALI with second and third order derivatives agree very well, only the latter are shown. Blue: MCMC-sampled likelihood for a 2-bin tomography analysis of the same data set without using a prior on $h$.}
\label{HSTprior}
\end{figure*}

As a complement to the parameter constraints in Sect.~\ref{sect_comparison}, we investigate which non-Gaussianities are still present if we combine the 2d-weak lensing survey with a Gaussian prior on $h$ with standard deviation $\sigma_h = 0.03$, roughly corresponding to the precision of current local constraints on the Hubble constant \citep{2011ApJ...730..119R}. Fig.~\ref{HSTprior} shows that even when using this prior, the posterior likelihood is not peaked sharply enough but non-Gaussianities remain. It also shows that switching to 2-bin tomography outperforms the inclusion of this prior into a 2d weak lensing analysis.

Figure \ref{Accurate} shows a comparison of the Fisher- and DALI-forecasts and MCMC for 2-bin tomography. The likelihood is then well approximated by a multivariate Gaussian. If one were to include more model parameters, e.g. a redshift dependent equation of state for dark energy, the likelihood would widen again, potentially leading to a non-Gaussian shape.

\begin{figure*}
\centering \includegraphics[width=\textwidth]{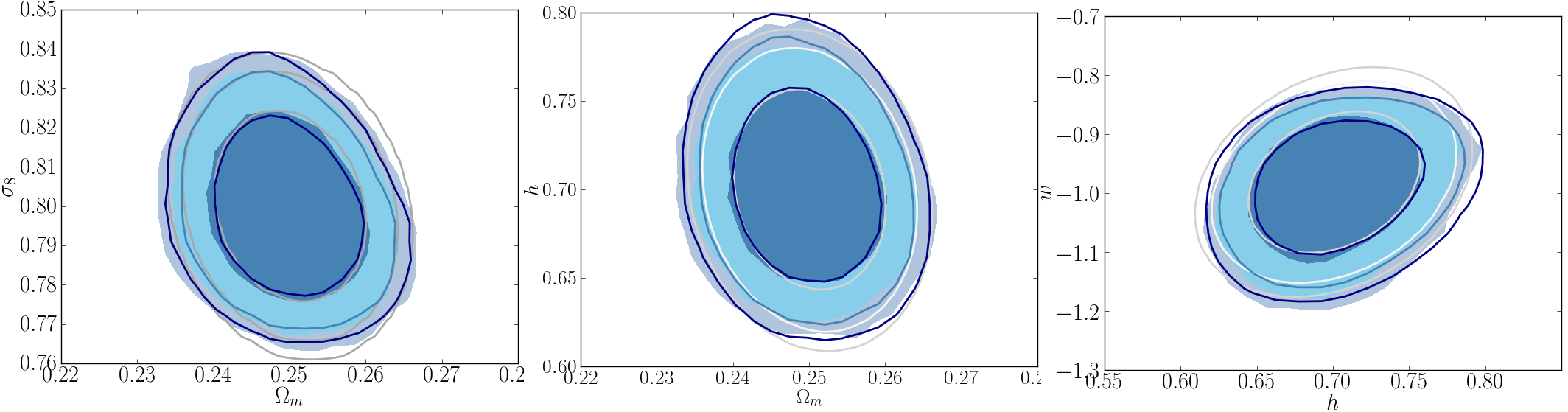} 
\caption{Two-bin tomography for the same weak lensing survey is able to break the degeneracies inherent in the parameter set, such that the overall constraints become much tighter and the Fisher matrix (grey contours) agree well with the MCMC-contours (solid blue). The MCMC-samples were generated with the Metropolis-Hastings algorithm. DALI finds nearly the same confidence contours as the Fisher matrix since only minor non-Gaussianity is present.}
\label{Accurate}
\end{figure*}

\section{Using DALI as approximate potential for Hamilton Monte Carlo}\label{sect_mcmc}
In order to have a comparison for DALI and the Fisher matrix, we generated MCMC-samples of the unapproximated likelihood. 

The Metropolis-Hastings algorithm \citep{1985LNP...240...62M} works well for approximately multivariate Gaussian likelihoods but has problems with following highly curved likelihoods. We therefore employed a Hamilton-Monte-Carlo (HMC) sampler, which uses Hamiltonian dynamics for describing a random walk on a potential $P$ corresponding to the logarithmic likelihood, 
\begin{equation}
 P(x_\mu) = - \ln(\mathcal{L}),
\label{potential}
\end{equation}
and a kinetic energy to introduce the randomness needed for sampling. 

The algorithm takes multiple leap frog steps along contours of approximately constant likelihood before performing a Metropolis-Hastings step by which it decides whether the new point is accepted or rejected \citep{Hamilton}. For each leap frog step, the HMC-sampler takes derivatives of the logarithmic likelihood and follows these, thereby adjusting well to curved likelihoods. 

Calculating derivatives of the true log-likelihood can be numerically costly. A gain in performance can then be achieved if the log-likelihood is replaced by an approximation which is fast to evaluate, such as DALI. Consequently, we do not use the log-likelihood of Eq.~(\ref{MCMC-likelihood}) for the sampler, but the DALI-approximation Eq.~(\ref{dalilike}) for the leap frog steps along the potential. Calculating the true weak-lensing likelihood is then only needed in the  Metropolis-Hastings steps. This procedure speeds up the performance of our sampler by a factor ranging between 30 and 100, depending on how many leap frog steps were done in each iteration of the HMC-algorithm.

\begin{figure}
\centering \includegraphics[width=0.5\textwidth]{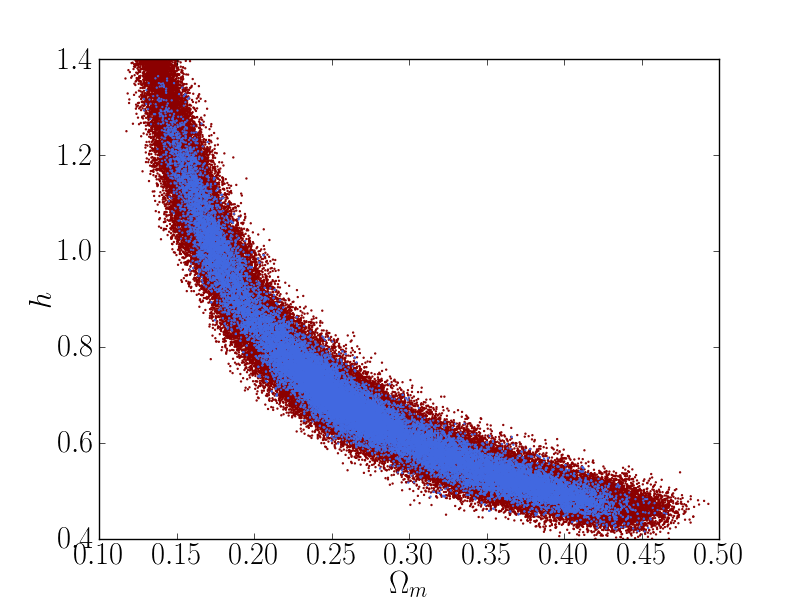} 
\caption{Samples from a (for plotting thinned) Hamilton Monte Carlo Chain: For the approximate potential, the tempered DALI-likelihood Eq.~(\ref{potential}) was used. Rejected samples are depicted in red, accepted samples are depicted in blue. Clearly visible is a red rim where accepted and rejected samples do not mix, demonstrating that the sampler was able to reach all points in parameter space that are preferred by the data. In the other two-dimensional planes, there also exists a ring of rejected samples.}
\label{MCMC}
\end{figure}

A potential issue with using DALI-contours to guide an HMC sampler is that DALI might exclude regions of the parameter space that are actually preferred by the true likelihood. In order to avoid this problem, we introduce a temperature to widen the potential Eq.~(\ref{potential}),
\begin{equation}
\ln P(x_\mu) \rightarrow \ln P(x_\mu)/T.
\label{potential_temp}
\end{equation}
If the temperature is set too high, the contours of the potential Eq.~(\ref{potential_temp}) will not generate samples that follow the true likelihood well. This leads to a reduction of the acceptance rate. We find that $T = 3$ leads to an acceptance rate between $0.3$ and $0.5$ while still giving the sampler the possibility to reach all regions in parameter space that are erroneously not covered by the DALI-approximation. In Fig.~\ref{MCMC} we plot samples of such an MCMC-chain, demonstrating that the sampler has indeed been able to cover the true likelihood fully: The accepted samples are surrounded by a rim of rejected samples. This rim shows that the sampler had the chance to explore regions of parameter space with low likelihood. 

In contrast, using the Fisher matrix as an approximate potential for the HMC-sampler has proven ineffective: since it does not capture the curvature of the likelihood, the sampler is frequently guided towards regions of extremely low likelihood if it follows the isocontours of the Fisher approximation. Consequently, even after adjusting the number of leap frog steps, no higher acceptance rate than  $0.02$ in our application could be gained, while many regions of the preferred parameter space were not sampled (in an acceptable time) at all.

% --- section: summary --- %
\section{Summary}\label{sect_summary}
In this paper, we have investigated different methods of how to treat parameter degeneracies when estimating the information content of a Euclid like weak lensing data set by calculating parameter constraints. The three possible ways are: $(i)$ break the degeneracies by using priors $(ii)$ include the parameter degeneracies by using a non-Gaussian likelihood and $(iii)$ break the degeneracies by substantially changing the data analysis, in this case changing from 2d-weak lensing to lensing tomography.

Additionally, we have demonstrated how the DALI-approximation allows to guide a Hamilton Monte Carlo sampler, such that highly curved likelihoods can be effectively sampled.
For 2d weak lensing we found strong non-Gaussianities in the likelihoods and:
\begin{itemize}
\item The Fisher matrix then extends far into the unphysical parameter range $\Omega_m < 0$ and $h < 0$, since the Gaussian approximation to the actual likelihood is not particularly good.
\item Introducing the priors $h > 0$ and $\Omega_m > 0$ makes the Fisher matrix approximated likelihood similar to the MCMC-likelihood. The Cramer-Rao inequality does not need to be fulfilled when unphysical parameter regions are covered by the Fisher matrix. 
\item The Fisher matrix predicts that 2d-weak lensing at Euclid precision will not be able to put a lower limit on the Hubble constant, whereas we find with MCMC-evaluations and DALI-approximations of the likelihood that $h>0.4$. 
\item The reason why the Fisher matrix fails are strong hyperbolic parameter degeneracies. Breaking these degeneracies by including priors masks that the actual weak lensing likelihood is able to measure the cosmological parameters including $h$ without the aid of external data sets.
\item We did not require the inclusion of any priors for DALI: the fact that it captures non-Gaussianities was sufficient for DALI being in agreement with the MCMC-sampled likelihood. 
\item Due to its inherent high symmetry, the Fisher matrix does not hint at the presence of dark energy as strongly, as an MCMC-evaluation or a DALI-approximation do: about a third of the Fisher matrix falls into parameter ranges of decelerated expansion, whereas about $90\%$ of the DALI- and MCMC-likelihood fall into the region of accelerated expansion.
\item The Figures of Merit (FoM) from the non-Gaussian DALI approximation are better in agreement with the FoM from MCMC, than the FoM from the Fisher matrix in the presence of strong non-Gaussianities: The DALI-FoM was at most $\sim 30\%$ larger than the MCMC-FoM, whereas the Fisher-FoM was up to $65\%$ too large.
\item Using DALI as an approximate potential for a Hamiltonian Monte Carlo sampler speeds up the sampling in a two-fold way: Firstly, all leapfrog steps become effectively numerically costless since evaluating the DALI likelihood is extremely fast. Secondly, as the DALI-contours already follow very well the isocontours of the real likelihood, the sampler is being guided towards relevant regions in parameter space. This increases the acceptance ratio. In our application, we were able to increase the acceptance rate from $0.02$ to $0.3 - 0.5$, meaning the sampler needs to try at least 15 times less samples in order to achieve the same number of accepted samples. Simultaneously, the evaluation time for each sample was on average cut down by a factor of about 80, since the leapfrog steps did not require the calculation of the real likelihood anymore.
\end{itemize}

Even after introducing a prior of local measurements of the Hubble constant into the 2d weak lensing analysis, non-Gaussianities remain in the combined likelihood. However, these are much less pronounced. We also found that tomographic weak lensing without a prior on $h$ leads to tighter parameter constraints than 2d weak lensing with a prior on $h$. For our five dimensional parameter set $\Omega_m,\sigma_8,n_s,h,w$, the likelihood for 2-bin tomography was already well Gaussian. Potentially, this might change if a redshift dependence of the dark energy equation of state is allowed.

% --- section: acknowledgements --- %
\section*{Acknowledgements}
We thank Luca Amendola for helpful discussions and support, and Martin Kilbinger for valuable comments. ES acknowledges financial support through the RTG \emph{'Particle Physics beyond the Standard Model'} and through the TRR33 project ``The Dark Universe'' of the German Science Foundation.

\appendix

% ---  --- %
\section{Improved third derivatives}
Already from the Fisher matrix approach it is known that numerical derivatives must be estimated accurately since they determine the extent and the orientation of the Fisher matrix. This issue also affects DALI since it also uses numerical derivatives.
In \dalipaper, DALI was applied to a data set of supernovae. There, the DALI-contours of Figs.~2c and~3c had been observed to leak out of the true likelihood shape. These plots had been generated with an old version of the DALI code which used a numerically fast but rough algorithm for calculating third derivatives: it took another derivative of precomputed and splined second derivatives. The new public version of DALI uses a slower but more accurate routine for calculating third derivatives: it calculates them by using finite differences on the original function, not any already derived quantities. Redoing the analysis of \dalipaper ~ with the improved code results in Fig.~\ref{Accurate}, demonstrating that with the more carefully conducted estimate of third derivatives, the erroneous leakage of the DALI-likelihoods disappears. 

\begin{figure}
\centering \includegraphics[width=0.5\textwidth]{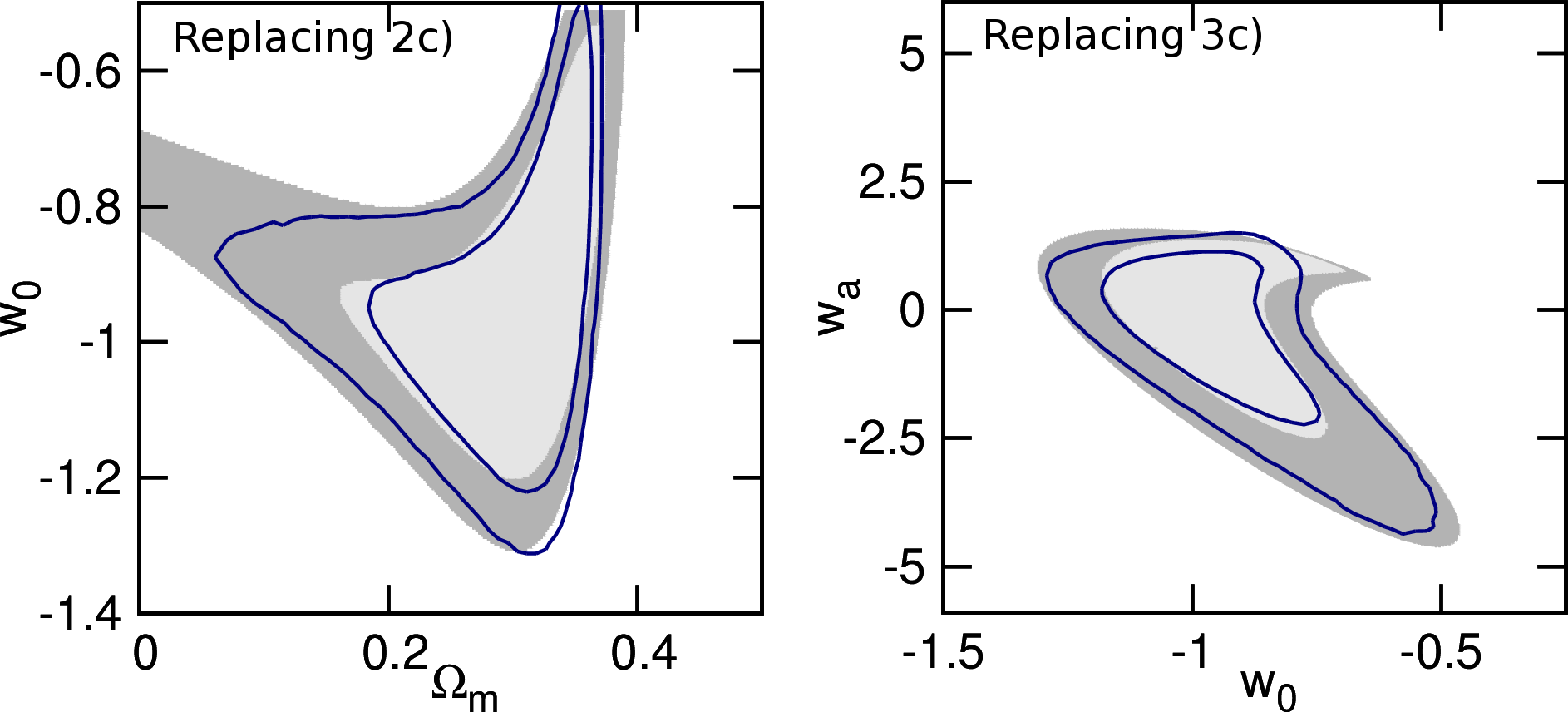} 
\caption{Replacement of the Fig.~2c and Fig.~3c in \citep{Sellentinetal}, using the new public version of DALI where third derivatives are calculated more accurately. The DALI-likelihood contours now follow the true likelihood better demonstrating that the observed mismatch between the true likelihood (grey) in \citep{Sellentinetal} and the DALI-contours was only due to numerically crude estimates.}
\label{Accurate}
\end{figure}

\section{The Cramer-Rao inequality and unphysical parameter ranges}
The Cramer-Rao inequality does not need to apply if the Fisher matrix covers unphysical parameter ranges. To illustrate this, we imagine a distribution function $f(\boX, \theta)$, where $\boX$ is the data set, and for simplicity only one parameter $\theta$ shall be estimated (else, one would simply need to marginalize over the other parameters).

The Cramer-Rao inequality actually holds for the Fisher-information $\mathcal{I}$, which is the averaged squared gradient of the log-likelihood,
\begin{equation}
 \mathcal{I} = \int f(\boX,\theta) [\partial_\theta \log(f(\boX,\theta)) ]^2 \mathrm{d}d.
\end{equation}
In contrast, the Fisher matrix is the averaged curvature of the negative log-likelihood
\begin{equation}
 F_{\theta\theta} = - \int f(\boX,\theta) \partial_\theta \partial_ \theta \log(f(\boX,\theta) ) \mathrm{d}d.
\label{curv}
\end{equation}

Explicitely calculating the second derivatives in Eq.~\ref{curv} shows that the Fisher matrix and the Fisher information are related by
\begin{equation}
F_{\theta\theta} = \mathcal{I} - \int \partial_\theta \partial_\theta f(\boX,\theta)\mathrm{d}d
\end{equation}
In order for the Fisher matrix to be identical to the Fisher information, the second term must vanish, which will be the case if the differentiation with respect to the parameters and the averaging over the data interchange. This however requires that the distribution $f(\boX,\theta)$ and its derivatives exist for all combinations of the data and the parameters and are finite. In the case of the Fisher matrix from Sect.~\ref{sect_comparison}, this is not fulfilled and the Cramer-Rao inequality then does not need to apply.

% --- section: bibliography --- %
\bibliography{bibtex/references}

\begin{thebibliography}{}

\bibitem[\protect\citeauthoryear{{Amara} \& {R{\'e}fr{\'e}gier}}{{Amara} \&
  {R{\'e}fr{\'e}gier}}{2007}]{2007MNRAS.381.1018A}
{Amara} A.,  {R{\'e}fr{\'e}gier} A.,  2007, MNRAS, 381, 1018

\bibitem[\protect\citeauthoryear{{Amendola} \& {Tsujikawa}}{{Amendola} \&
  {Tsujikawa}}{2010}]{Lucabook}
{Amendola} L.,  {Tsujikawa} S.,  2010, {Dark Energy: Theory and Observations}.
Cambridge University Press

\bibitem[\protect\citeauthoryear{{Bacon}, {Refregier} \& {Ellis}}{{Bacon}
  et~al.}{2000}]{2000MNRAS.318..625B}
{Bacon} D.~J.,  {Refregier} A.~R.,    {Ellis} R.~S.,  2000, MNRAS, 318, 625

\bibitem[\protect\citeauthoryear{{Bardeen}, {Bond}, {Kaiser} \&
  {Szalay}}{{Bardeen} et~al.}{1986}]{1986ApJ...304...15B}
{Bardeen} J.~M.,  {Bond} J.~R.,  {Kaiser} N.,    {Szalay} A.~S.,  1986,
  Astrophysical Journal, 304, 15

\bibitem[\protect\citeauthoryear{{Bartelmann}}{{Bartelmann}}{2010}]{2010CQGra..27w3001B}
{Bartelmann} M.,  2010, Classical and Quantum Gravity, 27, 233001

\bibitem[\protect\citeauthoryear{{Bartelmann} \& {Schneider}}{{Bartelmann} \&
  {Schneider}}{2001}]{2001PhR...340..291B}
{Bartelmann} M.,  {Schneider} P.,  2001, Physics Reports, 340, 291

\bibitem[\protect\citeauthoryear{{Castro}, {Heavens} \& {Kitching}}{{Castro}
  et~al.}{2005}]{2005PhRvD..72b3516C}
{Castro} P.~G.,  {Heavens} A.~F.,    {Kitching} T.~D.,  2005, Phys. Rev. D, 72,
  023516

\bibitem[\protect\citeauthoryear{{de Jong}, {Kuijken}, {Applegate}, {Begeman},
  {Belikov}, {Blake}, {Bout}, {Boxhoorn} \& et al.}{{de Jong}
  et~al.}{2013}]{2013Msngr.154...44J}
{de Jong} J.~T.~A.,  {Kuijken} K.,  {Applegate} D.,  {Begeman} K.,  {Belikov}
  A.,  {Blake} C.,  {Bout} J.,  {Boxhoorn} D.,    et al. 2013, The Messenger,
  154, 44

\bibitem[\protect\citeauthoryear{{Hajian}}{{Hajian}}{2007}]{Hamilton}
{Hajian} A.,  2007, Phys. Rev. D, 75, 083525

\bibitem[\protect\citeauthoryear{{Hannestad}, {Tu} \& {Wong}}{{Hannestad}
  et~al.}{2006}]{2006JCAP...06..025H}
{Hannestad} S.,  {Tu} H.,    {Wong} Y.~Y.,  2006, JCAP, 6, 25

\bibitem[\protect\citeauthoryear{{Heavens}}{{Heavens}}{2003}]{2003MNRAS.343.1327H}
{Heavens} A.,  2003, MNRAS, 343, 1327

\bibitem[\protect\citeauthoryear{{Heavens}, {Kitching} \& {Taylor}}{{Heavens}
  et~al.}{2006}]{2006MNRAS.373..105H}
{Heavens} A.~F.,  {Kitching} T.~D.,    {Taylor} A.~N.,  2006, Phys. Rev. D,
  373, 105

\bibitem[\protect\citeauthoryear{{Heymans}, {Grocutt}, {Heavens}, {Kilbinger},
  {Kitching}, {Simpson}, {Benjamin}, {Erben} \& {Hildebrandt}}{{Heymans}
  et~al.}{2013}]{2013MNRAS.432.2433H}
{Heymans} C.,  {Grocutt} E.,  {Heavens} A.,  {Kilbinger} M.,  {Kitching} T.~D.,
   {Simpson} F.,  {Benjamin} J.,  {Erben} T.,    {Hildebrandt} 2013, MNRAS,
  432, 2433

\bibitem[\protect\citeauthoryear{{Hu}}{{Hu}}{1999}]{1999ApJ...522L..21H}
{Hu} W.,  1999, ApJL, 522, L21

\bibitem[\protect\citeauthoryear{{Hu}}{{Hu}}{2002}]{2002PhRvD..66h3515H}
{Hu} W.,  2002, Phys. Rev. D, 66, 083515

\bibitem[\protect\citeauthoryear{{Huterer} \& {White}}{{Huterer} \&
  {White}}{2005}]{2005PhRvD..72d3002H}
{Huterer} D.,  {White} M.,  2005, Phys. Rev. D, 72, 043002

\bibitem[\protect\citeauthoryear{{Jain} \& {Taylor}}{{Jain} \&
  {Taylor}}{2003}]{2003PhRvL..91n1302J}
{Jain} B.,  {Taylor} A.,  2003, Physical Review Letters, 91, 141302

\bibitem[\protect\citeauthoryear{{Kaiser}, {Wilson} \& {Luppino}}{{Kaiser}
  et~al.}{2000}]{2000astro.ph..3338K}
{Kaiser} N.,  {Wilson} G.,    {Luppino} G.~A.,  2000, ArXiv e-prints 000338

\bibitem[\protect\citeauthoryear{{Kilbinger}, {Fu}, {Heymans}, {Simpson},
  {Benjamin} \& {Erben}}{{Kilbinger} et~al.}{2013}]{2013MNRAS.430.2200K}
{Kilbinger} M.,  {Fu} L.,  {Heymans} C.,  {Simpson} F.,  {Benjamin} J.,
  {Erben} 2013, MNRAS, 430, 2200

\bibitem[\protect\citeauthoryear{{Laureijs}, {Amiaux}, {Arduini},
  {Augu{\`e}res}, {Brinchmann}, {Cole}, {Cropper}, {Dabin}, {Duvet}, {Ealet} \&
  et al.}{{Laureijs} et~al.}{2011}]{EuclidStudyReport}
{Laureijs} R.,  {Amiaux} J.,  {Arduini} S.,  {Augu{\`e}res} J.~.,  {Brinchmann}
  J.,  {Cole} R.,  {Cropper} M.,  {Dabin} C.,  {Duvet} L.,  {Ealet} A.,    et
  al. 2011, ArXiv e-prints

\bibitem[\protect\citeauthoryear{{Linder} \& {Jenkins}}{{Linder} \&
  {Jenkins}}{2003}]{2003MNRAS.346..573L}
{Linder} E.~V.,  {Jenkins} A.,  2003, MNRAS, 346, 573

\bibitem[\protect\citeauthoryear{{Melchior}, {Suchyta}, {Huff}, {Hirsch},
  {Kacprzak}, {Rykoff}, {Gruen}, {Armstrong} \& et al.}{{Melchior}
  et~al.}{2015}]{2015MNRAS.449.2219M}
{Melchior} P.,  {Suchyta} E.,  {Huff} E.,  {Hirsch} M.,  {Kacprzak} T.,
  {Rykoff} E.,  {Gruen} D.,  {Armstrong} R.,    et al. 2015, MNRAS, 449, 2219

\bibitem[\protect\citeauthoryear{{Metropolis}}{{Metropolis}}{1985}]{1985LNP...240...62M}
{Metropolis} N.,  1985, in {Alcouffe} R.,  {Dautray} R.,  {Forster} A.,
  {Ledonois} G.,   {Mercier} B.,  eds, Lecture Notes in Physics, Berlin
  Springer Verlag Vol.~240 of Lecture Notes in Physics, Berlin Springer Verlag,
  {Monte-Carlo: In the Beginning and Some Great Expectations}.
p.~62

\bibitem[\protect\citeauthoryear{{Munshi}, {Smidt}, {Heavens}, {Coles} \&
  {Cooray}}{{Munshi} et~al.}{2010}]{2010arXiv1003.5003M}
{Munshi} D.,  {Smidt} J.,  {Heavens} A.,  {Coles} P.,    {Cooray} A.,  2010,
  ArXiv e-prints

\bibitem[\protect\citeauthoryear{{Refregier} \& {the DUNE
  collaboration}}{{Refregier} \& {the DUNE
  collaboration}}{2008}]{2008arXiv0802.2522R}
{Refregier} A.,  {the DUNE collaboration} 2008, ArXiv 0802.2522, 802

\bibitem[\protect\citeauthoryear{{Riess}, {Macri}, {Casertano}, {Lampeitl},
  {Ferguson}, {Filippenko}, {Jha}, {Li} \& {Chornock}}{{Riess}
  et~al.}{2011}]{2011ApJ...730..119R}
{Riess} A.~G.,  {Macri} L.,  {Casertano} S.,  {Lampeitl} H.,  {Ferguson} H.~C.,
   {Filippenko} A.~V.,  {Jha} S.~W.,  {Li} W.,    {Chornock} R.,  2011, ApJ,
  730, 119

\bibitem[\protect\citeauthoryear{{Sch{\"a}fer} \& {Heisenberg}}{{Sch{\"a}fer}
  \& {Heisenberg}}{2012}]{2012MNRAS.423.3445S}
{Sch{\"a}fer} B.~M.,  {Heisenberg} L.,  2012, MNRAS, 423, 3445

\bibitem[\protect\citeauthoryear{{Sellentin}}{{Sellentin}}{2015}]{Sellentin2015}
{Sellentin} E.,  2015, ArXiv e-prints

\bibitem[\protect\citeauthoryear{{Sellentin}, {Quartin} \&
  {Amendola}}{{Sellentin} et~al.}{2014}]{Sellentinetal}
{Sellentin} E.,  {Quartin} M.,    {Amendola} L.,  2014, MNRAS, 441, 1831

\bibitem[\protect\citeauthoryear{{Smith}, {Peacock}, {Jenkins}, {White},
  {Frenk}, {Pearce}, {Thomas}, {Efstathiou} \& {Couchman}}{{Smith}
  et~al.}{2003}]{2003MNRAS.341.1311S}
{Smith} R.~E.,  {Peacock} J.~A.,  {Jenkins} A.,  {White} S.~D.~M.,  {Frenk}
  C.~S.,  {Pearce} F.~R.,  {Thomas} P.~A.,  {Efstathiou} G.,    {Couchman}
  H.~M.~P.,  2003, MNRAS, 341, 1311

\bibitem[\protect\citeauthoryear{{Sugiyama}}{{Sugiyama}}{1995}]{1995ApJS..100..281S}
{Sugiyama} N.,  1995, Astrophysical Journal supplement, 100, 281

\bibitem[\protect\citeauthoryear{{Takada} \& {Jain}}{{Takada} \&
  {Jain}}{2004}]{2004MNRAS.348..897T}
{Takada} M.,  {Jain} B.,  2004, MNRAS, 348, 897

\bibitem[\protect\citeauthoryear{{Takada} \& {White}}{{Takada} \&
  {White}}{2004}]{2004ApJ...601L...1T}
{Takada} M.,  {White} M.,  2004, ApJL, 601, L1

\bibitem[\protect\citeauthoryear{{Tegmark}, {Taylor} \& {Heavens}}{{Tegmark}
  et~al.}{1997}]{1997ApJ...480...22T}
{Tegmark} M.,  {Taylor} A.~N.,    {Heavens} A.~F.,  1997, Astrophysical
  Journal, 480, 22

\bibitem[\protect\citeauthoryear{{Van Waerbeke}, {Mellier}, {Erben},
  {Cuillandre}, {Bernardeau}, {Maoli}, {Bertin}, {McCracken}, {Le F{\`e}vre},
  {Fort}, {Dantel-Fort}, {Jain} \& {Schneider}}{{Van Waerbeke}
  et~al.}{2000}]{2000A&A...358...30V}
{Van Waerbeke} L.,  {Mellier} Y.,  {Erben} T.,  {Cuillandre} J.~C.,
  {Bernardeau} F.,  {Maoli} R.,  {Bertin} E.,  {McCracken} H.~J.,  {Le
  F{\`e}vre} O.,  {Fort} B.,  {Dantel-Fort} M.,  {Jain} B.,    {Schneider} P.,
  2000, AAP, 358, 30

\bibitem[\protect\citeauthoryear{{Wittman}, {Tyson}, {Kirkman}, {Dell'Antonio}
  \& {Bernstein}}{{Wittman} et~al.}{2000}]{2000Natur.405..143W}
{Wittman} D.~M.,  {Tyson} J.~A.,  {Kirkman} D.,  {Dell'Antonio} I.,
  {Bernstein} G.,  2000, Nature, 405, 143

\end{thebibliography}
\bibliographystyle{mn2e}

\bsp

\label{lastpage}

\end{document}